\documentclass[prl,twocolumn,showpacs,nofootinbib]{revtex4-1}
  
\usepackage{amsmath}
\usepackage{nicefrac}
\usepackage{graphicx}
\usepackage{microtype}
\usepackage{placeins}
\usepackage{physics}
\usepackage{mdframed}
\usepackage{mathrsfs}
\usepackage{nicefrac}
\usepackage{multirow}

\usepackage{soul}
\usepackage[normalem]{ulem}

\usepackage[english]{babel}
\usepackage[applemac]{inputenc}
\usepackage[T1]{fontenc}

\newcommand{\beq}{\begin{equation}}
\newcommand{\eeq}{\end{equation}}
\newcommand{\beqa}{\begin{eqnarray}}
\newcommand{\eeqa}{\end{eqnarray}}

\newcommand\sgn{\operatorname{sgn}}

\usepackage{enumitem}
\setlist[description]{leftmargin=*}

\begin{document}

\title{Diagrammatic Monte Carlo approach to angular momentum\\[5pt] in quantum many-particle systems}
\author{G. Bighin$^{1}$, T. V. Tscherbul$^{2}$, M. Lemeshko$^{1}$}
\date{\today}

\affiliation{$^{1}$IST Austria (Institute of Science and Technology Austria), Am Campus 1, 3400 Klosterneuburg, Austria\\
$^{2}$Department of Physics, University of Nevada, Reno, NV, 89557, USA}

\begin{abstract}

We introduce a Diagrammatic Monte Carlo (DiagMC) approach to angular momentum properties of quantum many-particle systems possessing a  macroscopic number of degrees of freedom. The treatment is based on a diagrammatic expansion that merges the usual Feynman diagrams with the angular momentum diagrams known from atomic and nuclear structure theory, thereby incorporating the non-Abelian algebra inherent to quantum rotations. Our approach is applicable at arbitrary coupling, is free of systematic errors and of finite size effects, and naturally provides access to the impurity Green function. We exemplify the technique by obtaining an all-coupling solution of the angulon model, however, the method is quite general and can be applied to a broad variety of systems in which particles exchange quantum angular momentum with their many-body environment.

\end{abstract}

\maketitle


Quantum many-body systems involving coupled angular momenta are ubiquitous in physics. For example, in solids the electron spin and orbital angular momenta are coupled to the lattice degrees of freedom, which manifests itself in the Einstein-de Haas and Barnett effects~ \cite{Einstein:1915,Barnett:1915}. A complete understanding of the intricate many-body dynamics governing these effects, if achieved, would have far-reaching applications, from ultrafast control of magnetism \cite{Kirilyuk:2010,Mentink:2018} to spintronics~\cite{MatsuoPRL11} to solid-state quantum computing~\cite{Bassett:2014}. Among atomic and molecular systems, molecules immersed in superfluid helium \cite{Toennies:2004} or solid para-hydrogen \cite{Momose:1998}, as well as Rydberg atoms in Bose-Einstein condensates~\cite{Balewski:2013}, exchange their orbital angular momentum with the quantum many-body bath~\cite{Lemeshko:2016}. Similarly,  molecules spinning inside `cages' in perovskite crystals \cite{Lahnsteiner:2016} exchange angular momentum with the lattice, which was shown to influence  photovoltaic performance of solar-cell materials~\cite{Chen:2017}. Finally, since molecular reactivity depends on molecular orientation in space~\cite{Levine:2009}, modeling the effects of a fluctuating quantum environment on molecular rotations is a crucial step towards one of the main goals of chemistry -- controlling chemical reactions in solutions and on surfaces. Describing angular momentum dynamics in such many-body systems, however, represents a seemingly impossible task, for it requires addition of an \textit{essentially infinite} number of quantum angular momenta.

Numerical studies of the systems listed above usually follow `atomistic' approaches, such as path-integral \cite{Marx:1999,Cantano:2016,Zilich:2005} and diffusion \cite{Zilich:2004,Paesani:2001,Lee:1999} Monte Carlo, as well as density functional theory \cite{Hernando:2007fe,Ancilotto:2017}, where the many-body environment is modeled as a cluster of finite size in real space. These approaches make it challenging to reach the thermodynamic limit, are prone to systematic errors due to a discrete time step, while also providing only a limited insight into the angular momentum properties of the individual components of the many-particle system. The Diagrammatic Monte Carlo (DiagMC) technique~\cite{Prokofev:1998gza,Prokofev:1998,Mishchenko:2000co} represents an alternative approach: based on the idea of stochastic sampling in the space of Feynman diagrams, it works in continuous time and in the thermodynamic limit -- taking full advantage of the second quantization formalism. Through the years, DiagMC has established itself as a powerful and elegant method to obtain a numerically exact solution for a variety of many-particle problems \cite{Kulagin:2013a,Kulagin:2013b,Gukelberger:2014,Kozik:2010,Boninsegni:2006,VanHoucke:2013a,VanHoucke:2013b,Huang:2016,Deng:2015,Macridin:2004,VanHoucke:2012ic}, including the Fr\"ohlich polaron \cite{Prokofev:1998gza, Prokofev:1998,Mishchenko:2000co,Greitemann:2017uh}, the Holstein~\cite{Macridin:2003tu}, spinful~\cite{Mishchenko:2001ja} and Bose~\cite{Vlietinck:2015gt} polarons, excitons \cite{Mishchenko:2008gu}, and many-polaron ensembles \cite{Mishchenko:2014}. All applications of DiagMC to impurity problems to date, however, dealt either with structureless impurities (e.g.~spherical atoms) or impurities possessing a very simple internal structure (e.g.~spin-$\nicefrac{1}{2}$ systems or Jahn-Teller polarons \cite{Kornilovitch:2000}). An extension to particles possessing angular momentum degrees of freedom poses a substantial challenge due to the non-Abelian nature of  quantum rotations. Namely, a many-body state resulting from the coupling of $n$ angular momenta -- where $n$ can get arbitrarily large in the strong-coupling regime -- requires one to use $n$ Wigner $3j$-symbols, making it very difficult to go beyond the perturbative regime.

In this Letter, we develop a DiagMC approach to angular momentum properties of quantum many-body systems possessing a \textit{macrosopic} number of degrees of freedom, in equilibrium. We exemplify the capabilities of the approach by applying it to the angulon problem -- an extended impurity (e.g.\ a molecule) whose rotation is coupled to collective excitations of bosons. While maintaining the advantages of DiagMC -- working in the thermodynamic limit and continuous time allows to avoid finite-size effects and systematic errors -- our approach greatly extends its scope of application, establishing a far-reaching connection between DiagMC, molecular, and condensed-matter problems. The molecular Green's function and the angular momentum properties are naturally embedded in the formalism and easily accessed. Our approach differs in an essential way from the DiagMC treatment of other impurity problems, as we introduce a new class of updates which is necessary to explore the intricate diagram space of quantum rotations in a many-body environment.

The DiagMC approach described here is quite general, and can be applied to, in principle, any system whose orbital or rotational degrees of freedom are perturbed by a many-body environment. Without loss of generality, we exemplify the DiagMC technique by applying it to the angulon Hamiltonian, which can be seen as the rotational counterpart of the Fr\"ohlich polaron model \cite{Frohlich:1954}. The angulon model has been  originally derived for an ultracold molecule rotating in a weakly-interacting BEC \cite{Lemeshko:2016,Midya:2016un}. Later,  it was shown that angulons provide a phenomenological tool to describe molecules interacting with dense quantum solvents, such as superfluid $^4$He, in good agreement with experiment~\cite{Lemeshko:2017,Shepperson:2017gb,Cherepanov:2017cf}. In the case of a linear-rotor impurity immersed into a bath of bosons,  the angulon Hamiltonian reads \cite{Schmidt:2016du,Schmidt:2015hc,Lemeshko:2016}:
\beq
\hat{H} = B \mathbf{J}^2 +  \sum_{k \lambda \mu} \omega_{k} \hat{b}^\dagger_{k \lambda \mu} \hat{b}_{k \lambda \mu} +  \sum_{k \lambda \mu} U_\lambda(k) \left[ Y^*_{\lambda \mu} (\hat{\theta},\hat{\phi}) \hat{b}^\dagger_{k \lambda \mu} + \text{h.c.} \right],
\label{eq:hang}
\eeq
where $\sum_k \equiv \int \mathrm{d} k$ and units of $\hbar \equiv 1$ are used hereafter. The rotational constant, $B=1/(2I)$, is expressed in terms of the molecular moment of inertia $I$,  the bath is described by the field operators, $\hat{b}^\dagger_{k \lambda \mu}$ ($\hat{b}_{k \lambda \mu}$), which create (destroy) a bosonic excitation with linear momentum $k$, angular momentum $\lambda$, and angular momentum projection along the $z$ axis, $\mu$, and $\omega_k$ is the dispersion relation for the bosons. The third term of Eq.~\eqref{eq:hang} gives the impurity-bath interaction, where $Y_{\lambda \mu} (\hat \theta, \hat \phi)$ are the spherical harmonic operators~\cite{Varshalovich:1988} and $U_\lambda(k)$ is the angular-momentum dependent potential, which can be derived from the molecule-bath potential energy surface. The $\lambda$ summation of Eq.~(\ref{eq:hang}) is truncated at some cutoff, $n_\lambda$. Typically, only a few $\lambda$-channels are sufficient to describe the behavior of a rotating molecule in a many-body environment \cite{Lemeshko:2017}.

{\it Feynman rules for the angulon}. To solve the angulon quantum impurity problem, we introduce the retarded propagator for a free rotor, describing imaginary-time evolution between the initial and final angular configurations, $\Omega_i = \{\theta_i, \phi_i \}$ and $\Omega_f = \{\theta_f, \phi_f \}$ \cite{Bighin:2017jz}:
\beq
G_0 (\Omega_i, \Omega_f, \tau) = \sum_j \frac{2j + 1}{4 \pi} P_j(\cos \gamma_{if}) e^{-E_j \tau} \Theta(\tau),
\label{eq:g0def}
\eeq
where $\gamma_{if}$ is the angle between $\Omega_i$ and $\Omega_f$, $E_j = B j (j+1)$ is the free rotor energy, $\Theta(\tau)$ is Heaviside's step function, and $P_j$ are the Legendre polynomials. In a similar way, we introduce the retarded phonon propagator \cite{Bighin:2017jz}:
\beq
D (\Omega_i, \Omega_f, \tau) = \sum_{k \lambda} \frac{2\lambda + 1}{4 \pi} P_\lambda(\cos \gamma_{if}) |U_\lambda(k)|^2 e^{- \omega_k \tau} \Theta(\tau) \; 
\label{eq:ddef}
\eeq

Let us consider the imaginary-time evolution of the impurity+bath system from the initial state, $\ket{\Omega_i}_\text{imp} \otimes \ket{0}_\text{bos}$, to the final state, $\ket{\Omega_f}_\text{imp} \otimes \ket{0}_\text{bos}$, where $\ket{0}_\text{bos}$ represents the boson vacuum. For  this purpose, we   introduce the imaginary-time Green function, $G (\Omega_i, \Omega_f; \tau)$, which can be expressed through an infinite series expansion \cite{Bighin:2017jz}:
\beq
G (\Omega_i, \Omega_f ; \tau) = G_0 (\Omega_i, \Omega_f ; \tau) + \sum_{n=1}^{\infty} G^{(n)} (\Omega_i, \Omega_f ; \tau),
\label{eq:fullg}
\eeq
where
\begin{widetext}
\beq
\begin{split}
G^{(n)} (\Omega_i, \Omega_f ; \tau) = \sum_{\{p_i\}} \int \mathrm{d} \Omega_1 \ldots \mathrm{d} \Omega_{2n} \int \mathrm{d} \tau_1 \ldots \mathrm{d} \tau_{2n} \ G_0 (\Omega_i, \Omega_1, \tau_1) G_0 (\Omega_1, \Omega_2, \tau_2 - \tau_1) \ldots G_0 (\Omega_{2n},\Omega_f, \tau - \tau_{2n}) \times \\
\times D (\Omega_{p_1}, \Omega_{p_2}, \tau_{p_2} - \tau_{p_1}) \ldots D (\Omega_{p_{2n-1}}, \Omega_{p_{2n}}, \tau_{p_{2n}} - \tau_{p_{2n-1}})
\label{eq:gn}
\end{split}
\eeq
\end{widetext}
Here, the summation over $\{p_i\}$ covers all unordered permutations of all unordered pairs of the $2n$ time or angular variables appearing at order $n$.

\begin{figure}[t]
\centering
    \includegraphics[width=.995\linewidth]{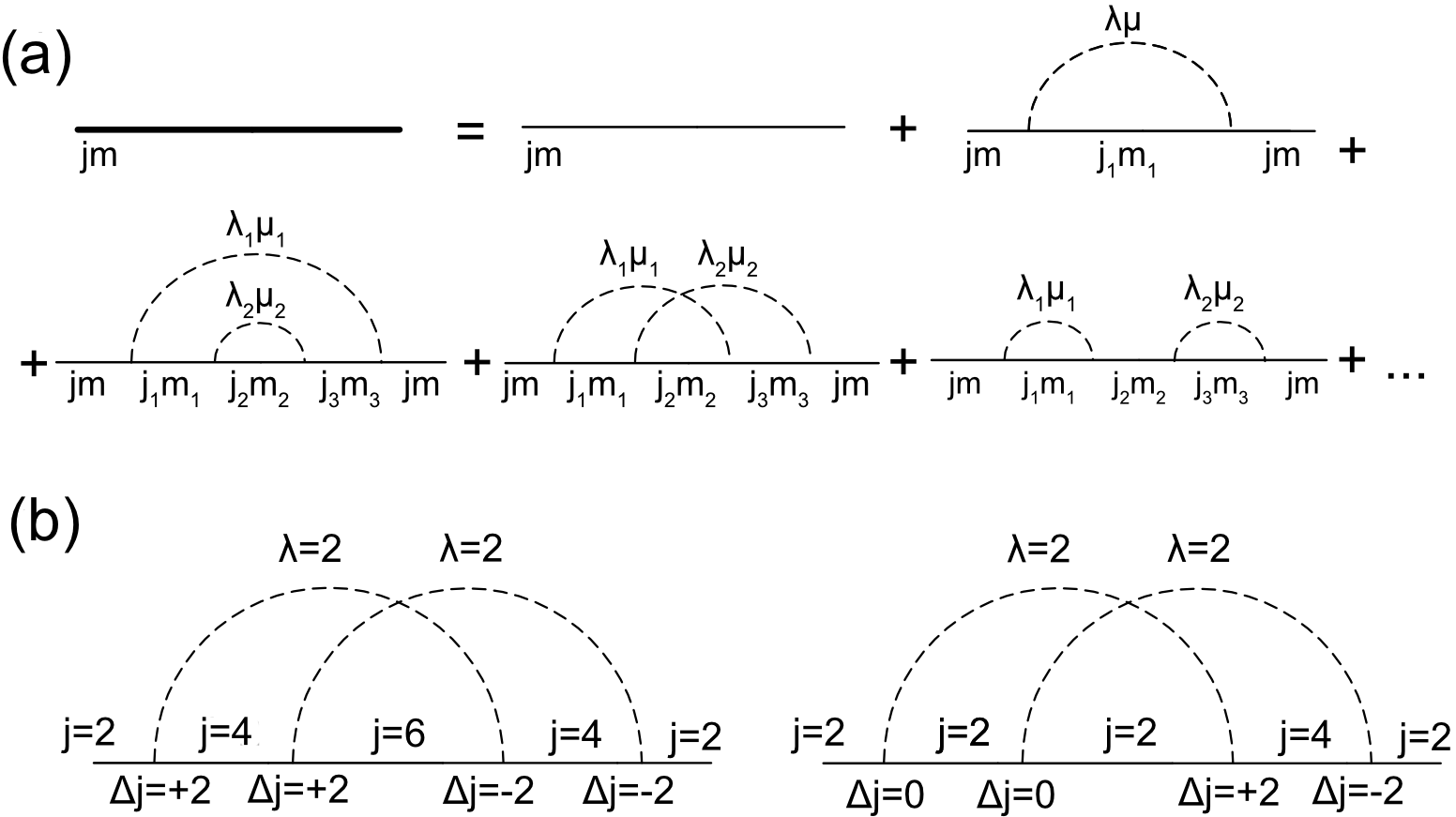}
\caption{(a) A few first terms in the diagrammatic expansion of the angulon Green function, $G_j (\tau)$. On each line, the $(jm)$ quantum numbers indicate the transferred angular momentum. (b) The diagram on the left can be generated using the `Change', `Add' and `Remove' updates, whereas that on the right is generated using the `Shuffle' update (there, angular momentum is not locally conserved within each phonon arc, i.e. $(\Delta j)_\text{start} \neq (\Delta j)_\text{end}$). The azimuthal quantum numbers $m$, $\mu$ are not shown.}
\label{fig:one}
\end{figure}

Equation (\ref{eq:gn}) has two crucial properties. First, since retarded propagators are used, it is completely equivalent to a time-ordered expression. Second, this equation  factorises when expressed in the angular momentum basis, due to the `spherical convolution' theorem \cite{sm3}. Therefore each term in the expansion \eqref{eq:gn} can be represented by a diagram resulting in a diagrammatic expansion similar -- at first glance -- to that for  the polaron \cite{Kholodenko:1983iu}. As opposed to the polaron problem, however, each line carries an angular momentum, $(j, m)$, rather than a linear momentum, $\mathbf{k}$, and each vertex has to enforce conservation of angular momentum~\cite{Bighin:2017jz}. A few lowest-order diagrams from the expansion are shown in Fig. \ref{fig:one}(a). The resulting diagrammatic rules associate to each rotor line the factor $(-1)^m G_{0,j}(\tau) = (-1)^m \exp(-B j (j+1) \tau)$ and to each phonon line the factor $(-1)^{\mu} D_{\lambda}(\tau) = (-1)^\mu \sum_{k} |U_\lambda (k)|^2 \exp(- \tau \ \omega_k)$. Each vertex, in turn, corresponds to
\begin{multline}
V^{\lambda \pm \mu}_{jm j' m'} = \sqrt{\frac{(2j+1) (2j'+1) (2\lambda+1)}{4 \pi}} \times \\
\times \begin{pmatrix}
j & \lambda & j' \\
- m & \pm \mu & m'
\end{pmatrix} \begin{pmatrix}
j & \lambda & j' \\
0 & 0 & 0
\end{pmatrix},
\label{eq:vertex}
\end{multline}
where we introduced  the Wigner $3j$-symbol \cite{Varshalovich:1988} and implied summation over all angular momenta and their $z$-projections associated with every rotor and phonon line, except for those corresponding to the initial and final free propagators. The $+$ ($-$) sign before $\mu$ indicates that the phonon line leaves (enters) the vertex (note that the orientation of each line is completely arbitrary as long as one line enters and leaves exactly one vertex). In what follows, Latin letters will be used for free propagators' quantum numbers, $(j_i, m_i)$, whereas Greek letters will denote the phonon propagators' quantum numbers, $(\lambda_j, \mu_j)$. We stress that the rules described above can be viewed as the rules of the graphical theory of angular momentum \cite{Yutsis:1962,Massot:1967hu,Wormer:2006ko,Balcar:2009,Judd:1963,Rudzikas:2007,Varshalovich:1988} dressed by a novel, many-body contribution due to the phonon bath.

{\it Diagrammatic Monte Carlo procedure}.  Each diagram in the   expansion of Eq.~(\ref{eq:fullg}) is composed of $N$ free propagators and $M$ phonon arcs, with $N=2M+1$, see Fig.~\ref{fig:one}(a). Each diagram  depends on a set of quantum numbers and  time variables, which can be represented as vectors: $\vec{L} = \{j_1, \ldots, j_N, \lambda_1, \ldots, \lambda_M\}$, $\vec{M} = \{m_1, \ldots, m_N, \mu_1, \ldots, \mu_M\}$, and $\vec{T} = \{ \tau_1, \ldots, \tau_N \}$.

The angulon Green function of Eq. (\ref{eq:fullg}) can be expanded in spherical harmonics~\cite{Bighin:2017jz, sm3}, which allows us to write it as a sum over all possible diagrams at any order, as well as  all the quantum numbers circulating on the internal lines:
\begin{equation}
G_j (\tau) = \sum_{n=0}^{\infty} \sum_{\xi_n} \sum_{\vec{L}, \vec{M}, \vec{T}} \mathcal{D}_\xi (\vec{L}, \vec{M},\vec{T},\tau),
\label{eq:gang0}
\end{equation}
(we omitted the $m$ index due to rotational invariance~\cite{sm3}),  
where $\xi$ is diagram topology at order $n$, and $ \mathcal{D}_{\xi_n} (\vec{T},\vec{L}, \vec{M},\tau)$ is the weight of a diagram with topology $\xi_n$, internal times $\vec{T}$, internal quantum numbers $\vec{L}$, $\vec{M}$, and total length $\tau$. This sum can be evaluated using DiagMC by designing a stochastic process sampling  the whole space of diagrams~\cite{Prokofev:1998gza,Prokofev:1998,Mishchenko:2000co}.

{\it Diagram updates}. We start by defining the `Change' update~\cite{Prokofev:1996,Prokofev:1998gza,Prokofev:1998,Mishchenko:2000co} which simply modifies the length of the last free propagator in a diagram.  The updated diagram length follows the distribution:
\beq
\mathcal{P}_\text{change} (\tau) = \frac{(E_j - \mu)e^{-(E_j - \mu) \tau}}{1-e^{-(E_j - \mu) \tau_\text{max}}},
\eeq
where $j$ is the angular momentum of the last propagator, and $\tau_\text{max}$ is the diagram length. The update has the acceptance ratio of unity, as the analogous update for the Fr\"ohlich polaron \cite{Prokofev:1998}.  

The `Add' update adds a new phonon arc to a diagram. We choose the angular momentum $\lambda$ circulating on the new phonon arc with uniform probability among the $n_\lambda$ allowed values, and the $z$-axis projection of angular momentum, $\mu = -\lambda \dots \lambda$, with uniform probability among the $2\lambda+1$ values. Then, a free propagator is chosen randomly among $n_\text{pr}$ free propagators. Let us label its initial (final) time as $\tau_1$ ($\tau_2$), with the  initial time for the new phonon arc, $\tau_\text{start}$, being chosen uniformly between $\tau_1$ and $\tau_2$. The final time is $\tau_\text{end} = \tau_\text{start} + \Delta \tau$, where $\Delta \tau$ is chosen according to a distribution with a normalized probability density,
\beq
W(\Delta \tau) \propto \sum_{k} |U_\lambda (k)|^2 \exp(- \Delta \tau \ \omega_k)
\eeq
Let us consider a diagram such as the ones shown in Fig.~\ref{fig:one}(a), and   define $\Delta j_i$ as the angular momentum difference between the $i$-th and the $(i+1)$-th free propagators, as shown in Fig.~\ref{fig:one}(b). Then for a newly-created vertex at $\tau_\text{start}$ ($\tau_\text{end}$) the corresponding difference is $\Delta j_\text{start}$ ($\Delta j_\text{end}$). It is straightforward to verify that the vertex rule of Eq.~(\ref{eq:vertex}) generally allows for more than one coupling of the free propagators under the new arc. More precisely, if we fix  all the phonon angular momenta and the angular momenta for propagators not enclosed by the new arc, $\Delta j_\text{start}$ and $\Delta j_\text{end}$ can take any values in the range of $-\lambda, -\lambda + 2, \ldots, \lambda -2, \lambda$, and similar conditions hold for the $\Delta j$'s in between. We choose $\Delta j_\text{start}$ with uniform probability between the $(\lambda + 1)$ possible values ($\Delta j$'s of all enclosed vertices are kept unchanged), and set $\Delta j_\text{end} = - \Delta j_\text{start}$ to enforce global angular momentum conservation (which also implies $- m + m' + \mu = 0$ at every vertex). The resulting probability for adding a specific phonon line is then:
\beq
\mathcal{P}_\text{add} = \frac{1}{n_\text{pr}} \frac{1}{n_\lambda} \frac{1}{2 \lambda + 1} \frac{1}{|\tau_{\text{end}} - \tau_{\text{start}}|} \frac{1}{\lambda + 1} W(\Delta \tau)
\label{eq:padd}
\eeq
The symmetric `Remove' update simply selects a phonon line among $n_\text{ph}$ possible candidates and removes it, therefore, the associated probability is
\beq
\mathcal{P}_\text{remove} = \frac{1}{n_\text{ph}} \; .
\label{eq:premove}
\eeq
The acceptance ratios for the `Add' and `Remove' updates are straightforwardly 
derived from $\mathcal{P}_\text{add}$ and $\mathcal{P}_\text{remove}$ and the diagram weights \cite{Prokofev:1998}.

Crucially, unlike in the case of the Fr\"ohlich polaron, the three updates described above do not span the entire configuration space: the `Add' and `Remove' updates generate only a subset of all the diagrams where for each phonon arc $\Delta j_\text{end} = - \Delta j_\text{start}$. However, there can be diagrams for which this condition is not satisfied.
 That is, while angular momentum conservation holds globally (i.e. the first and last impurity propagators of a diagram always carry the same angular momentum), a single phonon arc may have $\Delta j_\text{start} \neq - \Delta j_\text{end}$, i.e. it may subtract from the impurity a different number of quanta of angular momentum than it gives back. In every physical diagram, this local violation of angular momentum conservation, as shown in Fig. \ref{fig:one}(b), must be counterbalanced by an opposite local violation in one or several other phonon arcs, such that $\sum_i \Delta j_i = 0$ holds for the whole diagram.

It can be shown \cite{sm3} that if we divide a diagram in 1-particle-irreducible subdiagrams, within each subdiagram one must have $\sum_i \Delta j_i = 0$ (diagrams not respecting this constraint have zero weight). Then, in order to visit every diagram, we need to introduce another, `Shuffle' update, which selects one random 1-particle-irreducible subdiagram and changes the $\Delta j$ configuration within that cluster to another allowed one. A suitable practical approach consists in choosing a random $\Delta j$ value for each vertex in the subdiagram, and   starting  over if the constraint above is not met (typically, only a few iterations are required). The probability of selecting another configuration is an involved combinatorial problem on its own, however, since the update is clearly balanced with itself, one never needs to calculate the probability $\mathcal{P}_\text{shuffle}$, and the acceptance ratio will be simply given by the ratio of diagram weights.

Finally, we note that the balance requirement for the `Add'/`Remove' updates requires that the latter will remove only phonon lines for which $\Delta j_\text{end} = - \Delta j_\text{start}$. Therefore, $n_\text{ph}$ in Eq. (\ref{eq:premove}) should be interpreted as  the number of \textit{available} phonons.

\begin{figure}[t]
\centering
\includegraphics[width=\linewidth]{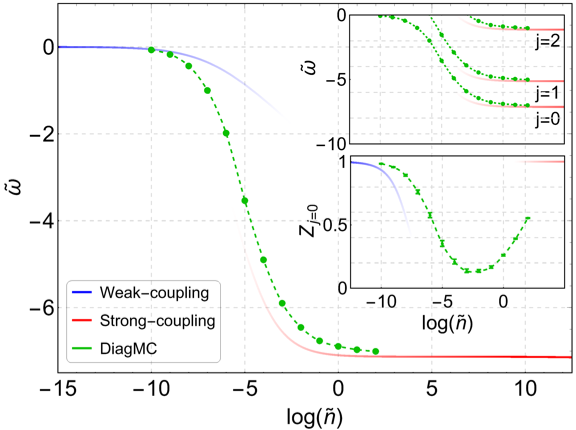}
\caption{The ground-state energy of the angulon Hamiltonian obtained using DiagMC (green circles) as a function of the dimensionless bath density, $\tilde{n}$, in comparison with the weak-coupling theory of Ref.~\cite{Schmidt:2015hc} (blue) and the strong-coupling theory of Ref.~\cite{Schmidt:2016du} (red). Upper inset: energy for the $j=0,1,2$ states obtained using DiagMC, in comparison with the  strong-coupling theory. Lower inset: quasiparticle weight for the $j=0$ state, as compared with the weak- and strong-coupling theories.}
\label{fig:two}
\end{figure}

{\it Negative weights}. Within the diagrammatic rules  introduced above, some of the diagrams can have a negative weight. This problem can be solved by sampling with respect to the absolute value of the diagram weight \cite{Gull:2013tg}, i.e. for a generic quantity $A$,
\beq
\langle A \rangle = \frac{\langle A \sgn \mathcal{D}_\xi \rangle_{|\mathcal{D}_\xi|}}{\langle \sgn \mathcal{D}_\xi \rangle_{|\mathcal{D}_\xi|}}
\label{eq:sign}
\eeq
We note, however, that this is \textit{not} the sign problem arising due to the fermionic anti-commutation relations -- here negative sign of some diagrams arises as the result of angular momentum coupling via $3j$-symbols. Indeed, we verified that the expectation value of the sign never falls below $0.6$ over the wide range of couplings we consider, implying that no nearly complete cancellation happens and that the error of sampled quantities does not increase exponentially.

{\it Numerical results}. In order to demonstrate the scope of the approach, we applied it to the ground- and excited-state energies of the angulon Hamiltonian, and compared the outcome of DiagMC with the weak- \cite{Schmidt:2015hc} and strong-coupling \cite{Schmidt:2016du} theories. Let us take the bath dispersion relation of the Bogoliubov form, $\omega_k = \sqrt{\epsilon_k ( \epsilon_k + 2 g_\text{bb} n)}$, with $g_\text{bb}=4 \pi a_\text{bb}/m$, the boson-boson scattering length  $a_{\text{bb}}=3.3 (m B)^{-\nicefrac{1}{2}}$, and $\epsilon_k=k^2/2m$. Following Refs.~\cite{Schmidt:2015hc,Schmidt:2016du} we define the angular-momentum dependent potential as $U_\lambda(k) = u_\lambda \sqrt{\frac{8 n k^2 \epsilon_k}{\omega_k (2 \lambda+1)}} \int \mathrm{d} r r^2 f_\lambda(r) j_\lambda(kr)$, 
with $n$ the density of the bosonic bath, and $j_\lambda$ the spherical Bessel function. We use Gaussian form factors, $f_\lambda(r)=(2 \pi)^{-\nicefrac{3}{2}} \exp[-r^2/(2 r_\lambda^2)]$ with interaction strength, $u_0 = 3.33 u_1 = 300 B$, and range, $r_0 = r_1 = 1.5 (m B)^{-\nicefrac{1}{2}}$, implying a $\lambda$-cutoff $n_\lambda=1$ in Eq. (\ref{eq:hang}).

The Green function in imaginary time can be accessed through a DiagMC sampling of Eq. (\ref{eq:gang0}), using the updates we introduced. The quasiparticle weight $Z_j$ and the energy of each state $E_j$ are obtained by fitting the long imaginary-time tail with $G_j(\tau) = Z_j \exp(-E_j \ \tau)$, where $E_j$ is the energy of the state with angular momentum $j$ and $Z_j$ is its quasiparticle weight. In Fig. \ref{fig:two} we present the results of DiagMC  for the energies of the ground state and of a few excited angulon states, as well as for the ground-state quasiparticle weight, showing that they are in good agreement with the analytic expressions in the limits of weak~\cite{Schmidt:2015hc} and strong~\cite{Schmidt:2016du} coupling. In the intermediate-coupling regime, which has not been characterized before, the ground-state energy interpolates smoothly between the weak and strong-coupling limits. We note that while higher excited states cannot be accessed in this parameter regime due to the stability condition of the perturbative expansion, $E_j < \min_k \omega_k$  \cite{Mishchenko:2000co},  they can be  studied by means of an analytic continuation of the imaginary-time results~\cite{Mishchenko:2000co,Bertaina:2017cd,Jarrell:1996is,Gubernatis:1991hi,Silver:1990fc}.

In summary, we have introduced a numerically exact approach to  quantum many-body systems involving coupled angular momenta. Based on the second quantization formalism, the present treatment is considerably less expensive compared to most atomistic approaches, and is free from systematic errors due to finite-size effects or time discretization. An essential new feature of our approach, as compared to DiagMC treatments of the Fr\"ohlich polaron \cite{Prokofev:1998,Mishchenko:2000co}, is the presence of a new `Shuffle' update which is necessary to explore the intricate diagram space of a rotating impurity.

While the study of real-time dynamics of the angulon Hamiltonian is a formidable task beyond the scope of this work, our approach lays the foundations for developing a real-time perturbation theory along the Keldysh contour {\cite{Schiro:2009ft}}, which would enable studying, e.g., the dynamics of molecules in quantum solvents \cite{Shepperson:2017gb} if used in conjunction with recent techniques devised to treat the dynamical sign problem~\cite{Cohen:2015}. Furthermore, a quantum impurity solver is a key ingredient of dynamical mean-field theory used to reveal the properties of  strongly-correlated condensed-matter systems~\cite{Georges:1996zz}. The technique described here can potentially be implemented as a building block of such numerical techniques with applications to collective molecular processes. 

Finally, it is important to emphasize that we used the angulon Hamiltonian merely as a prototypical example of a many-body system with angular momentum degrees of freedom. The DiagMC technique is quite general and can be straightforwardly extended to describe more complicated molecular geometries \cite{Cherepanov:2017cf}, to interactions terms different from the linear coupling of Eq.~(\ref{eq:hang}), or to translationally moving orbital impurities.

\begin{acknowledgments}
We acknowledge stimulating discussions with Kun Chen, Cesare Franchini, Thomas Hahn, Jacques Tempere, Nikolai Prokof'ev, and Boris Svistunov at various stages of this work. This work was supported by the Austrian Science Fund (FWF), project Nr. P29902-N27. 
\end{acknowledgments}

\clearpage
\pagebreak
\widetext
\begin{center}
\textbf{\large Supplemental Material:\\[5pt] Diagrammatic Monte Carlo approach to angular momentum\\[3pt] in quantum many-particle systems}
\end{center}
\setcounter{equation}{0}
\setcounter{figure}{0}
\setcounter{table}{0}
\setcounter{page}{1}
\makeatletter
\renewcommand{\theequation}{S\arabic{equation}}
\renewcommand{\thefigure}{S\arabic{figure}}
\renewcommand{\bibnumfmt}[1]{[S#1]}
\renewcommand{\citenumfont}[1]{S#1}

\section{Green's functions in the angular momentum basis}

We start from the Green function for a free quantum rotor~\cite{Dickhoff:2005,Favro:1960ie}
\beq
G_0(\Omega,\Omega'; t )= \sum_n \psi_n(\Omega) \psi_n^*(\Omega') e^{-\mathrm{i} E_n t},
\label{eq:appg0begin}
\eeq
where the index $n$ runs over all energy eigenstates  of the rotor, $E_n$, with $\psi_n$ being the corresponding eigenfunction. For a linear rotor the wavefunctions $\psi_n$ are the spherical harmonics, $n \equiv \{ j,m \}$, and $E_n  \equiv  E_j = B j (j+1)$, leading to:
\beq
G_0 (\Omega,\Omega' ; t) = \sum_{j m} Y^{*}_{j m} (\Omega) Y_{j m} (\Omega') e^{-\mathrm{i} E_j t} \; .
\eeq
We carry out the sum over $\mu$ by means of the spherical harmonics addition theorem
\beq
G_0 (\Omega,\Omega' ; t) = \sum_j \frac{2 j + 1}{4 \pi} P_j (\cos \gamma(\Omega,\Omega')) e^{-\mathrm{i} E_j t},
\eeq
where $\gamma(\Omega,\Omega')$ is the angle between $\Omega$ and $\Omega'$, and $P_j (x)$ are the Legendre polynomials. Finally, switching to imaginary time, one gets the rotor propagator,
\beq
G_0 (\Omega,\Omega' ; \tau) = \sum_j \frac{2 j + 1}{4 \pi} P_j (\cos \gamma(\Omega,\Omega')) e^{-E_j \tau},
\eeq
and -- in a completely analogous way~\cite{Bighin:2017jz} -- the phonon propagator,
\beq
D (\Omega_i, \Omega_f, \tau) = \sum_{k \lambda} \frac{2\lambda + 1}{4 \pi} P_\lambda(\cos \gamma(\Omega,\Omega')) |U_\lambda(k)|^2 e^{-\omega_k \tau}.
\eeq

\section{Spherical harmonics expansion}

Let us consider a function $f(\Omega_1, \Omega_2)$, where  $\Omega_1 = \left\{ \theta_1, \phi_1 \right\}$ and $\Omega_2 = \left\{ \theta_2, \phi_2 \right\}$. We define the spherical harmonics expansion of $f$ as
\beq
f_{j m j' m'} = \int \mathrm{d} \Omega_1 \mathrm{d} \Omega_2 \ Y^*_{j m}(\Omega_1) Y_{j' m'}(\Omega_2) \ f (\Omega_1, \Omega_2)
\label{eq:g1a}
\eeq
Assuming that $f$ depends only on the angle between its arguments, i.e.
\beq
f(\Omega_1,\Omega_2) = f(\gamma(\Omega_1,\Omega_2)),
\label{eq:rotinvar}
\eeq
where $\gamma(\Omega_1,\Omega_2) = \cos(\theta_1) \cos(\theta_2) + \sin(\theta_1) \sin(\theta_2) \cos(\phi_2 - \phi_1)$, one can show that $f_{j m j' m'}$  has the following structure~\cite{Bighin:2017jz}:
\beq
f_{j m j' m'} = f_j \delta_{j j'} \delta_{m,m'},
\label{eq:qed}
\eeq
where
\beq
f_j = 2 \pi \int_{-1}^1 \mathrm{d} x \ P_j (x) f(cos \gamma = x).
\eeq
Given that $f_j$ completely defines the spherical harmonics expansion of $f(\Omega_1,\Omega_2)$, with a slight abuse of terminology, we call $f_j$ the spherical harmonics expansion coefficients of $f$. It is worth stressing that due to rotational invariance of the problem all the Green functions introduced in the main text possess the property defined in Eq (\ref{eq:rotinvar}).

\section{The convolution theorem in the angular momentum basis}

We introduce the `spherical convolution' of $f(\Omega_1, \Omega_2)$ and $g(\Omega_1, \Omega_2)$, defined as
\beq
h(\Omega_1, \Omega_2) = \int \mathrm{d} \Omega' f(\Omega_1,\Omega') g(\Omega',\Omega_2) \;.
\eeq
Let us assume that $f$ and $g$ are functions of the angle between their arguments only, i.e. they satisfy the condition of Eq. (\ref{eq:rotinvar}). Due to rotational invariance their `spherical convolution' $h$ must have the same property as well. In Ref. \cite{Bighin:2017jz} we have shown that the spherical harmonics expansion coefficient $h_l$ of $h(\Omega_1, \Omega_2)$ has the following form:
\beq
h_j = f_j \ g_j,
\label{eq:hjfjgj}
\eeq
thereby extending the usual convolution theorem for the Fourier transform to the case of the spherical harmonics expansion. Introducing the $\star$ notation for the convolution in the spherical basis, one can then rewrite Eq. (\ref{eq:hjfjgj}) as
\beq
(f \star g)_{j} = f_j \ g_j,
\eeq
and extend the convolution theorem  to an arbitrary number of functions:
\beq
(f_1 \star f_2 \star \ldots \star f_n)_{j} = \prod_{i=1}^n (f_i)_j  \; .
\eeq

\section{A constraint on the one-particle-irreducible components}

Applying the Feynman rules as described in the main text, one can immediately see that the weight associated to a diagram is
\beq
\mathcal{D}_{\xi_n} (\vec{T}, \vec{L}, \vec{M}, \tau) = \prod_i^N (-1)^{m_i} G_{0,j_i} (\tau_i - \tau_{i-1}) \prod_j^M (-1)^{\mu_j} D_{\lambda_j} (\tau_{g(j)} - \tau_{h(j)}) \prod_k^{N-1} V^{\lambda_{f(k)} \sigma_k \mu_{f(k)}}_{j_k m_k j_{k+1} m_{k+1}}
\label{eq:prods}
\eeq
where the function $f(k)$ determines which phonon line is connected to the $k$-th vertex, effectively encoding the diagram topology $\xi_n$. Similarly $g(j)$ ($h(j)$) corresponds to the starting (ending) vertex of the $j$-th phonon line. Finally, $\sigma_k$ is $+1$ $(-1)$ if the a phonon line leaves (enters) the $k$-th vertex. As noted in the main text, the orientation of a line is never explicitly drawn on a diagram, since it is completely irrelevant, provided that each line enters exactly one vertex and exits exactly one vertex.

\begin{figure}[h!]
\centering
    \includegraphics[width=.60\linewidth]{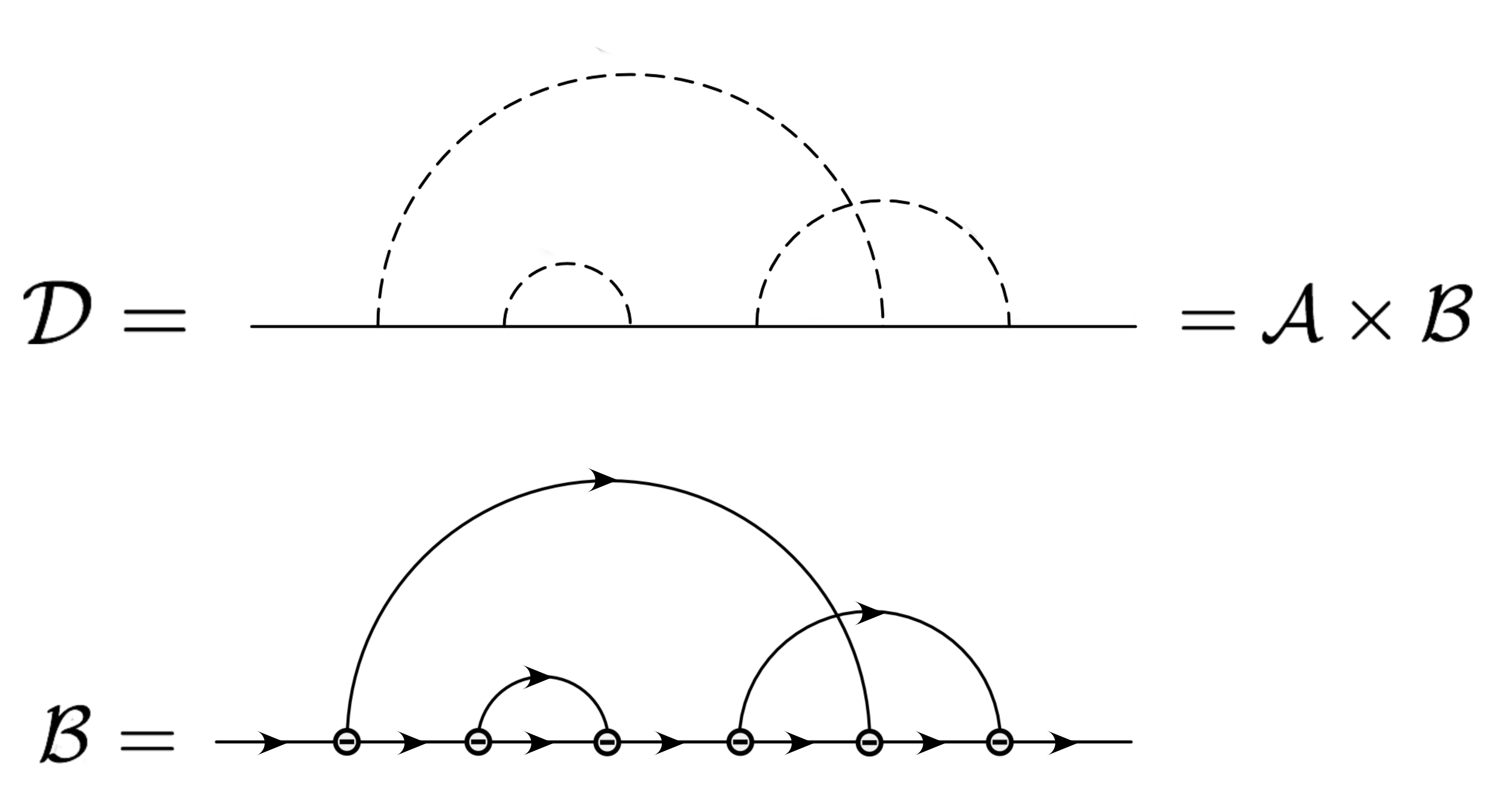}
\caption{Each diagram can be decomposed into two contributions: a `many-body' one, $\mathcal{A}$, and a `geometric' one, $\mathcal{B}$. The latter, depicted in the lower panel using the formalism of the graphical theory of angular momentum  \cite{Yutsis:1962,Massot:1967hu,Wormer:2006ko,Balcar:2009,Judd:1963,Rudzikas:2007,Varshalovich:1988}, can be calculated exactly by repeatedly applying the rules of Fig. \ref{fig:rules}.}
\label{fig:eq}
\end{figure}

\begin{figure}[bh]
\centering
    \includegraphics[width=.65\linewidth]{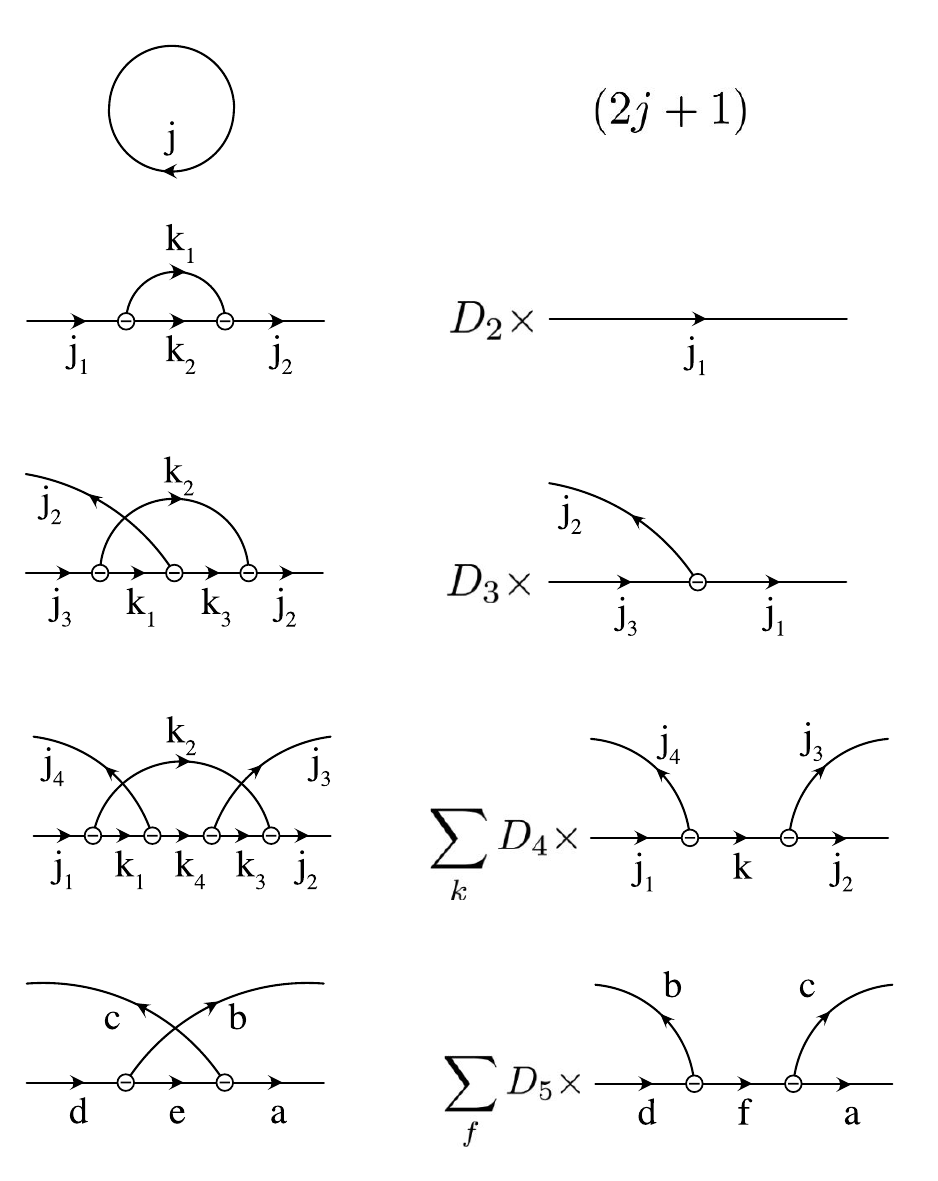}
\caption{The rules used to simplifying the angular momentum diagrams: each diagram on the left can be mapped to a simpler  diagram  on the right, provided that a coefficient, $D_i$,  and (in some cases) an additional summation, $\sum_{k}$, is introduced.}
\label{fig:rules}
\end{figure}

Now, the weight $\mathcal{D}_{\xi_n} (\vec{T}, \vec{L}, \vec{M}, \tau)$ corresponding to each diagram can be decomposed in two parts, following the reasoning we introduced in Ref.~\cite{Bighin:2017jz}:
\begin{equation}
\mathcal{D}_\xi (\vec{L}, \vec{M}) = \mathcal{A}_\xi (\vec{L}) \mathcal{B}_\xi (\vec{L}, \vec{M}) \; .
\label{eq:dab}
\end{equation}
The first part, $\mathcal{A}$,  encapsulates  the many-body aspects of the problem, whereas the second part, $\mathcal{B}$, contains the geometrical information related to the coupling of the angular momenta circulating in the diagram. Throughout the present Section, the time dependence of the diagram weights, $\mathcal{D}$, and of each propagator will be omitted for the sake of simplicity. From Eq.~(\ref{eq:prods}), one can easily see that a possible choice for $\mathcal{A}$ and $\mathcal{B}$ is
\beq
\mathcal{A}_\xi (\vec{L}) = \prod_{i}^{N} (-1)^{j_i} G_{0,j_i} \prod_{j}^{M} (-1)^{\lambda_j} D_{\lambda_j} \prod_k^{N-1} \sqrt{\frac{(2j_k+1) (2j_{k+1}+1) (2\lambda_{f(k)}+1)}{4 \pi}} \begin{pmatrix}
j_k & \lambda_{f(k)} & j_{k+1} \\
0 & 0 & 0
\end{pmatrix}
 \; ,
\eeq
and
\beq
\mathcal{B}_\xi (\vec{L},\vec{M}) = (-1)^{\sum_i l_i - m_i + \sum_j \lambda_j - \mu_j}  \prod_{k}^{N-1} \begin{pmatrix}
l_k & \lambda_{f(k)} & l_{k+1} \\
- m_k & \sigma_k \mu_{f(k)} & m_{k+1}
\end{pmatrix} \; .
\eeq
The decomposition of Eq. (\ref{eq:dab}) has been chosen such that $\mathcal{B}$ has an immediate interpretation in terms of the graphical theory of angular momentum \cite{Yutsis:1962,Massot:1967hu,Wormer:2006ko,Balcar:2009,Judd:1963,Rudzikas:2007,Varshalovich:1988}, as shown in Fig. \ref{fig:eq}.
Namely, the analytical expression $\mathcal{B}$ corresponds to an angular momentum diagram with the same topology as the corresponding Feynman diagram, $\mathcal{D}$, if one replaces each vertex with a `$-$' node and  orients each line from left to right, using the notations and conventions of Ref. \cite{Massot:1967hu}. This opens up the possibility of a systematic simplification of the sums over $\vec{M}$, for instance by implementing the graphical-algebraic techniques of Refs.~\cite{Yutsis:1962,BarShalom:1988in,Fack:1997fp,VanDyck:2003gl}. Repeated application of the rules of Fig. \ref{fig:rules} can simplify  diagrams with arbitrary complexity, at the expense of introducing additional numerical factors or summations, contained in each $D_i$, which depend  on $\vec{L}$ only. The numerical factors from Fig. \ref{fig:rules} read:

\begin{align}
D_2 &= (-1)^{j_1+k_1+k_2} \frac{\delta_{j1,j2}}{2 j_1 + 1}, \\
D_3 &= (-1)^{k_1+j_2+k_3} \begin{Bmatrix} j_1 & j_2 & j_3 \\
k_1 & k_2 & k_3
\end{Bmatrix}, \\
D_4 &= (-1)^{k+k_1+k_2+k_3+k_4+j_3+j_4} (2k+1) \begin{Bmatrix} j_1 & j_4 & k \\
k_4 & k_2 & k_1
\end{Bmatrix}  \begin{Bmatrix} j_2 & j_3 & k \\
k_4 & k_2 & k_3
\end{Bmatrix}, \\ 
D_5 &= (-1)^{b+c+e+f} (2f+1) \begin{Bmatrix} a & b & f \\
d & c & e
\end{Bmatrix},
\end{align}
and a systematic application of the rules allows one to completely carry out the sums over $\vec{M}$, leading to:
\begin{equation}
\sum_{\vec{M}} \mathcal{D}_\xi (\vec{L}, \vec{M}) = \mathcal{D}_\xi ( \vec{L} ) \; .
\label{eq:sumb}
\end{equation}
We verified that after carrying out the sums over $\vec{M}$ as in Eq. (\ref{eq:sumb}), each weight $\mathcal{D}_\xi ( \vec{L} )$ is positive. This means that the sum of Eq. (7) of the main text can be rewritten as a sum over $n$, $\xi_n$, $\vec{L}$ and $\vec{T}$ involving only positive weights. However, this approach is not advantageous, since the the computation of a diagram weight $\mathcal{D}_\xi ( \vec{L} )$ through Eq.~\eqref{eq:sumb} can be very expensive, in particular at higher orders. Much better results are achieved through a stochastic sampling of all quantum numbers, including $\vec{M}$.

Nonetheless, there is an important application of the rules introduced in the present Section. Let us consider a diagram, $\mathcal{D}$, with more than one one-particle-irreducible component. Of course, the corresponding angular momentum diagram, $\mathcal{B}$, will have the same topology and, henceforth, the same number of components. By repeatedly applying the rules of Fig. \ref{fig:rules} one can simplify one of the components to a bubble, and then contract the bubble to a single line by means of the second rule of Fig. \ref{fig:rules}. This effectively removes one one-particle-irreducible component at the expense of introducing a numerical prefactor, which, however, might be quite complex. 

At the last step -- when a bubble is contracted to a single line -- the Kronecker $\delta$ in $D_2$ requires that the angular momentum $j_1$ circulating \textit{before} the bubble must be the same as the angular momentum $j_2$ circulating \textit{after} the bubble. Going all the way back to the original angular momentum diagram, $\mathcal{B}$, before the simplification process, this means that if a one-particle-irreducible component does not start and end with the same angular momentum $j$, then $\mathcal{B}$ is be zero. This implies that, within a one-particle-irreducible component of a diagram, the condition
\beq
\sum_i \Delta j_i = 0
\eeq
must always hold, otherwise the weight $\mathcal{B}$ and, consequently, the diagram weight $\mathcal{D}$ would vanish.

\end{document}